\documentclass[10pt,twocolumn,a4paper]{IEEEtran}
\usepackage{graphicx}
\usepackage{caption,color}
\usepackage{subcaption}
\usepackage{amsthm,amssymb}
\usepackage{amsmath}
\usepackage{algorithm}
\usepackage{algorithmic}

\begin{document}

\title{\huge Study of Opportunistic Cooperation Techniques using Jamming and Relays for Physical-Layer Security in Buffer-aided Relay Networks}

\author{Xiaotao Lu and Rodrigo C. de~Lamare
\thanks{Xiaotao Lu is with the Communications Research Group, Department of Electronics, University of York, YO10 5DD York, U.K., R. C. de Lamare is with CETUC, PUC-Rio, Brazil and with the
Communications Research Group, Department of Electronics, University of York,
YO10 5DD York, U.K. This work was supported by CNPq and FAPERJ. E-mails:
xtl503@york.ac.uk; rodrigo.delamare@york.ac.uk.}}

\maketitle
\pagenumbering{gobble}
\begin{abstract}

In this paper, we investigate opportunistic relay and jammer
cooperation schemes in multiple-input multiple-output (MIMO)
buffer-aided relay networks. The network consists of one source, an
arbitrary number of relay nodes, legitimate users and eavesdroppers,
with the constraints of physical layer security. We propose an
algorithm to select a set of relay nodes to enhance the legitimate
users' transmission and another set of relay nodes to perform
jamming of the eavesdroppers. With Inter-Relay interference (IRI)
taken into account, interference cancellation can be implemented to
assist the transmission of the legitimate users. Secondly, IRI can
also be used to further increase the level of harm of the jamming
signal to the eavesdroppers. By exploiting the fact that the jamming
signal can be stored at the relay nodes, we also propose a hybrid
algorithm to set a signal-to-interference and noise ratio (SINR)
threshold at the node to determine the type of signal stored at the
relay node. With this separation, the signals with high SINR are
delivered to the users as conventional relay systems and the low
SINR performance signals are stored as potential jamming signals.
Simulation results show that the proposed techniques obtain a
significant improvement in secrecy rate over previously reported
algorithms.

\end{abstract}


\section{Introduction}

In broadcast channels, secure transmission is difficult to achieve due to the
broadcast nature of wireless communication systems. Traditional encryption
techniques are implemented in the network layer. With complex algorithms,
encryption keys which are nearly unbreakable are generated to ensure security
while their costs are extremely high. To reduce the costs of encryption
algorithms, researchers are investigating novel security techniques in the
physical layer of wireless systems. Physical-layer security has been first
illustrated in \cite{Shannon} from the viewpoint of information theory. The
feasibility of physical-layer security has been discussed by Shannon at the
theoretical level in the paper. Later on in \cite{Wyner} a wire-tap channel
which can achieve positive secrecy rate has been proposed by Wyner under the
assumption that the users experience a better channel than eavesdroppers. Since
then, the wire-tap model along with other techniques such as the broadcast
channel \cite{Csiszar}, MIMO channels \cite{Oggier}, \cite{Tie}, artificial
noise \cite{Goel}, beamforming \cite{Junwei} as well as relay systems
\cite{Oohama} have been studied. In this work, we focus on multiple-antenna m
relay systems and cooperation schemes
\cite{mmimo,wence,Costa,delamare_ieeproc,TDS_clarke,TDS_2,switch_int,switch_mc,smce,TongW,jpais_iet,TARMO,keke1,kekecl,keke2,Tomlinson,dopeg_cl,peg_bf_iswcs,gqcpeg,peg_bf_cl,Harashima,mbthpc,zuthp,rmbthp,Hochwald,BDVP},\cite{delamare_mber,rontogiannis,delamare_itic,stspadf,choi,stbcccm,FL11,jio_mimo,peng_twc,spa,spa2,jio_mimo,P.Li,jingjing,did,bfidd,mbdf}.
involving relay and jamming strategies in relays equipped with buffers.

\subsection{Prior and Related Work}

In recent years \cite{Mukherjee}, the concept of physical-layer security with
multiuser wireless networks has been investigated in numerous studies.
Approaches to achieving physical-layer security include the design of transmit
precoding strategies without the need for a secret key and the exploitation of
the wireless communication medium to develop secret keys over public channels
\cite{scharf,bar-ness,pados99,reed98,hua,goldstein,santos,qian,delamarespl07,xutsa,delamaretsp,kwak,xu&liu,delamareccm,wcccm,delamareelb,jidf,delamarecl,delamaresp,delamaretvt,jioel,delamarespl07,delamare_ccmmswf,jidf_echo,delamaretvt10,delamaretvt2011ST,delamare10,fa10,lei09,ccmavf,lei10,jio_ccm,ccmavf,stap_jio,zhaocheng,zhaocheng2,arh_eusipco,arh_taes,dfjio,rdrab,dcg_conf,dcg,dce,drr_conf,dta_conf1,dta_conf2,dta_ls,song,wljio,barc,jiomber,saalt}.
Relay-based cooperative systems \cite{Mukherjee} are an important evolution of
secure transmission strategies that can further improve the performance of
wireless systems. In this context, buffer-aided relay nodes have recently drawn
much attention \cite{Zlatanov}, \cite{Gaojie} and \cite{Huang} due to their
potential to further improve the secrecy rate in wireless transmissions as
compared to standard relays. In our prior investigation with \cite{Keke1},
\cite{Keke2} and \cite{Keke3} we have reported precoding techniques
\cite{Xiaotao1} as well as buffer-aided relay system \cite{Xiaotao2} which
improve the secrecy rate performance in multiuser MIMO systems.

Prior work on buffer-aided relay systems with secure transmissions
has been considered in half-duplex and full-duplex systems. In
\cite{Zlatanov}, the system model is described as one source, one
half-duplex decode-and-forward (DF) buffer relay and one
destination. Regarding the availability of the channel state
information at the transmitter (CSIT), two schemes with fixed-rate
transmission and mixed-rate transmission have been proposed. Based
on the instantaneous signal-to-noise ratio (SNR) of the source-relay
and relay-destination links a solution to the throughput-optimal
problem has been reported \cite{Zlatanov}. 
In \cite{Gaojie}, a max-ratio relay selection policy has been proposed to
optimize the secrecy transmission rate with the consideration of exact and
average gains of the eavesdroppers's channels. In \cite{Huang}, a two-hop
half-duplex buffer-aided relay system has been studied, where a relay selection
which adapts reception and transmission time slots on the channel quality is
proposed and the selection parameters are optimized to maximize the secrecy
throughput or minimize the secrecy outage probability (SOP). Half-duplex
systems can avoid the interference for the transmission at the relay nodes and
their drawback lies in the requirement of two or more time slots for
transmission, which decreases the transmission rate significantly.

Compared with half-duplex systems, full-duplex systems can provide a better
performance in terms of transmission rate. An opportunistic relay scheme has
been applied to buffer-aided systems in \cite{Nomikos1}, \cite{Nomikos2} and
\cite{Lee}. In the opportunistic relay scheme, the inter-relay interference
(IRI) is an important aspect that should be taken into account. In
\cite{Nomikos1}, IRI cancellation has been combined for the first time with
buffer-aided relays and power adaptation to mitigate IRI and minimize the
energy consumption. With one source and one destination, a new relay selection
policy has been analyzed in terms of outage probability and diversity.
Furthermore, in \cite{Nomikos2} a distributed joint relay-pair selection has
been proposed with the aim of rate maximization in each time slot. By exploring
the feasibility of IRI cancellation at the relay nodes, it gives the threshold
to avoid increased relay-pair switching and CSI acquisition. With relay
selection, in \cite{Lee} and \cite{Jingchao} jammer selection as well as a
joint relay and jammer selection have been proposed. In such systems, relays
may obtain a better transmission rate for the legitimate users, whereas jammers
can interfere in the transmission to the eavesdropper, resulting in improvement
of the secrecy rate. There are very few works in the literature which consider
cooperation between opportunistic buffer-aided relay schemes with jamming
techniques.

\subsection{Contributions}

Our work focuses on the use of the interference among buffer-aided
relays to assist the secure transmission to the users. The major
contributions in our paper are:
\begin{itemize}

  \item Novel multi-user relay systems with relay and jammer function selection to achieve high secrecy rates.

  \item Novel multi-user buffer-aided relay systems with signals stored in the buffers which are capable of
  jamming eavesdroppers to achieve high secrecy rate.

  \item Secrecy rate analyses of the proposed relay and jammer
  selection schemes
\end{itemize}

The rest of this paper is organized as follows. In Section II, the
system model and the performance metrics are introduced. A brief
review of the buffer relay selection is included in Section III. The
proposed relay and jammer function selection (RJF) as well as the
buffer-aided RJF (BF-RJF) selection are introduced in Section IV. In
Section IV, we discuss and present the simulation results. The
conclusions are given in Section V.

\subsection{Notation}
Bold uppercase letters ${\boldsymbol A}\in {\mathbf{C}}^{M\times N}$
denote matrices with size ${M\times N}$ and bold lowercase letters
${\boldsymbol a}\in {\mathbf{C}}^{M\times 1}$ denote column vectors
with length $M$. Conjugate, transpose, and conjugate transpose are
represented by $(\cdot)^\ast$, $(\cdot)^T$ and $(\cdot)^H$
respectively; $\boldsymbol I_{M}$ is the identity matrix of size
$M\times M$; $\rm diag \{\boldsymbol a\}$ denotes a diagonal matrix
with the elements of the vector $\boldsymbol a$ along its diagonal;
$\mathcal{CN}(0,\sigma_{n}^{2})$ represents complex Gaussian random
variables with $i.i.d$ entries with zero mean and $\sigma_{n}^{2}$
variance.

\section{System Model and Performance Metrics}

In this section, we introduce the buffer-aided relay system model
and describe the data transmission. The problem statement is then
presented along with the performance metrics used to assess the
proposed and existing techniques.

\subsection{System Model}

\begin{center}
\begin{figure}[h]
\centering
\includegraphics[scale=0.6]{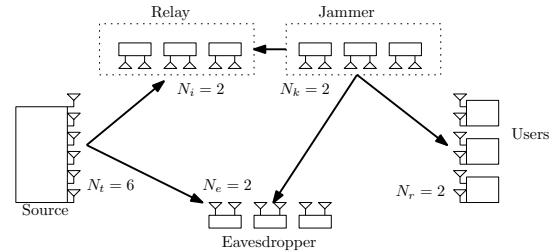}
\caption{System model of a MU-MIMO system with $M$ users, $N$ eavesdroppers $T$ relays and $K$ jammers}
\label{fig:sys}
\end{figure}
\end{center}

Fig. \ref{fig:sys} gives a description of a transmitter with $N_{t}$ antennas
used to transmit the data streams to $M$ users along with $N$ eavesdroppers.
With $T$ relays and $K$ jammers, in each time slot the selected relays and the
eavesdroppers receive the signal from both the source and the jammers. Note
that the signals from the jammers are the signals transmitted in the previous
time slots. They exploit the buffers at the relay nodes. Each relay and jammer
is equipped with $N_{i}$ and $N_{k}$ antennas. At the receiver side each user
and eavesdropper is equipped with $N_{r}$ and $N_{e}$ receive antennas. In this
system, we assume that the eavesdroppers do not jam the transmission of each
user, relay or jammer and that the eavesdropper channel is a flat-fading MIMO
channel. The quantities ${\boldsymbol H}_{i}\in {\mathbf{C}}^{N_{i}\times
N_{t}}$ and ${\boldsymbol H}_{e}\in {\mathbf{C}}^{N_{e}\times N_{t}}$ denote
the channel matrix from the source directly to the ith relay and the eth
eavesdropper, respectively. The quantities ${\boldsymbol H}_{ke}\in
{\mathbf{C}}^{N_{e}\times N_{k}}$ and ${\boldsymbol H}_{kr}\in
{\mathbf{C}}^{N_{r}\times N_{k}}$ denote the channel matrix of the eth
eavesdropper to the kth jammer and rth user to the kth jammer, respectively.
The channel between the kth relay to the ith relay is described by
${\boldsymbol H}_{ki}\in {\mathbb{C}}^{N_{i}\times N_{k}}$.

The vector ${\boldsymbol s}_{r}^{(t)}\in {\mathbb{C}}^{N_{r}\times
1}$ represents the data symbols to be transmitted corresponding to
each user in time slot $t$. The total transmit signal at the
transmitter can be expressed as ${\boldsymbol s}^{(t)}={\left[
{{\boldsymbol s}_{1}^{(t)}}^{T} \quad {{\boldsymbol
s}_{2}^{(t)}}^{T} \quad {{\boldsymbol s}_{3}^{(t)}}^{T} \quad \cdots
\quad {{\boldsymbol s}_M^{(t)}}^{T}\right]}^{T} $. In each phase the
received signal ${\boldsymbol y}_{i}^{(t)}\in
{\mathbb{C}}^{N_{i}\times 1}$ at each relay node can be expressed as
\begin{equation}
{\boldsymbol y}_{i}^{(t)}={\boldsymbol H}_{i}{\boldsymbol s}^{(t)}
+\sum_{k=1}^{K}{\boldsymbol H}_{ki}{\boldsymbol
y}_{k}^{(pt)}+\boldsymbol{n}_{i}, \label{eqn:yit}
\end{equation}
In (\ref{eqn:yit}), the superscript $(\cdot)^{(pt)}$ represents the previous
time slot when the signal is stored as a jamming signal in the buffer at the
relay nodes. The second term ${\boldsymbol H}_{ik}{\boldsymbol y}_{k}^{(pt)}$
is regarded as the inter-relay interference (IRI) between the ith relay and the
kth relay. The received signal ${\boldsymbol y}_{k}^{(pt)}$ is determined as
the jamming signal according to a signal-to-interference-plus-noise ratio
(SINR) criterion illustrated in the paper. With the theorem in \cite{Nomikos1},
IRI can be eliminated. The received signal at the eth eavesdropper is given by
\begin{equation}
{\boldsymbol y}_{e}^{(t)}={\boldsymbol H}_{e}{\boldsymbol s}^{(t)}
+\sum_{k=1}^{K}{\boldsymbol H}_{ke}{\boldsymbol
y}_{k}^{(pt)}+\boldsymbol{n}_{e}, \label{eqn:yet}
\end{equation}
where for the eavesdropper, the second term ${\boldsymbol
H}_{ek}{\boldsymbol y}_{k}^{(pt)}$ acts as the jamming signal and
this jamming signal can not be removed without the knowledge of the
channel from the kth jammer to the eth eavesdropper.

In (\ref{eqn:yit}) and (\ref{eqn:yet}), the IRI term between relay nodes or the
jamming signal to the eavesdropper is also the transmit signal from the relays
nodes to the destination. We assume that the transmit signal from the relay
nodes is described by ${\boldsymbol r}^{(t)}={\left[ {{\boldsymbol
y}_{1}^{(pt_{1})}}^{T} \quad {{\boldsymbol y}_{2}^{(pt_{2})}}^{T} \quad
{{\boldsymbol y}_{3}^{(pt_{3})}}^{T} \quad \cdots \quad {{\boldsymbol
y}_{T}^{(pt_{T})}}^{T}\right]}^{T}$. Note that the superscript $(\cdot)^{(pt)}$
represents the previous time slot due to the nature of relay nodes with
buffers. The values for the $i$th relay node can be different so the previous
time slot is represented by $(\cdot)^{(pt_{i})}$. The received signal at the
destination is expressed as
\begin{equation}
{\boldsymbol y}_{r}^{(t)}= \sum_{k=1}^{T}{\boldsymbol H}_{kr}{\boldsymbol
y}_{k}^{(pt_{k})}+\boldsymbol{n}_{r}, \label{eqn:yrt}
\end{equation}
Depending on the cancellation of the IRI at the relay nodes, two
types of schemes can be applied. Without IRI cancellation, based on
(\ref{eqn:yit}) the SINR at relay node $i$ is given by
\begin{equation}
\varGamma_{i}^{(t)}=\dfrac{\boldsymbol{\gamma}_{S,R_{i}}}{\varphi(k,i)\boldsymbol{\gamma}_{R_{k},R_{i}}+N_{i}\sigma_{i}^{2}},
\label{eqn:SINRiwoc}
\end{equation}
where $\varphi(k,i)$ is the factor indicating the performance
of IRI cancellation which we will discuss later and $\boldsymbol{\gamma}_{m,n}$ represents the
instantaneous received signal power for the link
$m\longrightarrow n$:
\begin{equation}
\boldsymbol{\gamma}_{S,R_{i}}=\mathrm {trace} ( \boldsymbol{H}_{i}\boldsymbol{H}_{i}^{H}),
\label{eqn:gammasr}
\end{equation}
\begin{equation}
\boldsymbol{\gamma}_{R_{k},R_{i}}
=\mathrm {trace} (\boldsymbol{H}_{ki}\boldsymbol{H}_{i}^{(pt)}{\boldsymbol{H}_{i}^{(pt)}}^{H}\boldsymbol{H}_{ki}^{H}),
\label{eqn:gammaki}
\end{equation}
Note that the superscript $\boldsymbol{H}_{i}^{(pt)}$ is applied due to the
nature of the signal stored in the buffers. The SINR at the eavesdropper node e
$\varGamma_{e}^{(t)}$ as well as the legitimate user r $\varGamma_{r}^{(t)}$
can be expressed as
\begin{equation}
\varGamma_{e}^{(t)}=\dfrac{\boldsymbol{\gamma}_{S,E_{e}}}{\boldsymbol{\gamma}_{R_{k},E_{e}}+N_{e}\sigma_{e}^{2}},
\label{eqn:SINRewoc}
\end{equation}
and
\begin{equation}
\varGamma_{r}^{(t)}=\dfrac{\boldsymbol{\gamma}_{R_{k},R_{r}}}{N_{r}\sigma_{r}^{2}},
\label{eqn:SINRrwoc}
\end{equation}
where in (\ref{eqn:SINRewoc}) and (\ref{eqn:SINRrwoc}) we have
\begin{equation}
\boldsymbol{\gamma}_{S,E_{e}}=\mathrm {trace} (\boldsymbol{H}_{e}\boldsymbol{H}_{e}^{H}),
\label{eqn:gammase}
\end{equation}
\begin{equation}
\boldsymbol{\gamma}_{R_{k},E_{e}}=\mathrm {trace} (\boldsymbol{H}_{ke}
\boldsymbol{H}_{i}^{(pt)}{\boldsymbol{H}_{i}^{(pt)}}^{H}\boldsymbol{H}_{ke}^{H}),
\label{eqn:gammare}
\end{equation}
\begin{equation}
\boldsymbol{\gamma}_{R_{k},R_{r}}=\mathrm {trace} (
\boldsymbol{H}_{kr}\boldsymbol{H}_{i}^{(pt)}
{\boldsymbol{H}_{i}^{(pt)}}^{H}\boldsymbol{H}_{kr}^{H}),
\label{eqn:gammakr}
\end{equation}
If the IRI cancellation can be performed at the relay node,
$\varphi(k,i)\boldsymbol{\gamma}_{R_{k},R_{i}}=0$ and the SINR at relay node i,
eavesdropper e and receiver r can be expressed, respectively, as
\begin{equation}
\varGamma_{i}^{(t)}=\dfrac{\boldsymbol{\gamma}_{S,R_{i}}}{N_{i}\sigma_{i}^{2}},
\label{eqn:SINRiwc}
\end{equation}
\begin{equation}
\varGamma_{e}^{(t)}=\dfrac{\boldsymbol{\gamma}_{S,E_{e}}}{\boldsymbol{\gamma}_{R_{k},E_{e}}+N_{e}\sigma_{e}^{2}},
\label{eqn:SINRewc}
\end{equation}
\begin{equation}
\varGamma_{r}^{(t)}=\dfrac{\boldsymbol{\gamma}_{R_{k},R_{r}}}{N_{r}\sigma_{r}^{2}},
\label{eqn:SINRrwc}
\end{equation}
According to \cite{Nomikos1}, the feasibility of IRI cancellation in
single-input-single-output (SISO) channels is considered. When we consider the
MIMO scenario with the same definition the factor $\varphi(k,i)$ can be
expressed as,
\begin{equation*}
\varphi(k,i)=
\begin{cases}
0& \text{if $\det \Big(({{\frac{P}{N_{t}}}\boldsymbol{H}_{i} \boldsymbol{H}_{i}^{H}+\boldsymbol{I}})^{-1}{{\frac{P}{N_{k}}}\boldsymbol{H}_{ki}\boldsymbol{H}_{ki}^{H}}\Big)\geqslant \gamma_{0}$}\\
1 &\text{otherwise},
\end{cases}
\label{eqn:varphi}
\end{equation*}
When interference cancellation is feasible, $\varphi(k,i)=0$, the
interfering signal is firstly decoded and then subtracted at the
relay prior to the decoding of the source signal. In this case, the
received signal at the relay node is not affected by the IRI.

\subsection{Problem Formulation}

In this subsection, we present the problem formulation and describe
the main performance metrics used to assess the performance of the
proposed algorithms.

The MIMO system secrecy capacity without consideration of interference is expressed as \cite{Tie}:
\begin{equation}
\begin{split}
C_{s} &=\max_{{\boldsymbol Q}_{s}\geq 0, \rm Tr({\boldsymbol Q}_{s}) =
E_{s}}\log(\det({\boldsymbol I}+{\boldsymbol H}_{ba} {\boldsymbol Q}_{s} {\boldsymbol H}_{ba}^H))\\
& \quad -\log(\det({\boldsymbol I}+ {\boldsymbol H}_{ea}{\boldsymbol
Q}_{s} {\boldsymbol H}_{ea}^H)),
\end{split}
\label{eqn:Rs1}
\end{equation}
In (\ref{eqn:Rs1}) ${\boldsymbol Q}_{s}$ is the covariance matrix
associated with the signal and ${\boldsymbol H}_{ba}$ and
${\boldsymbol H}_{ea}$ represent the links between the source to the
users and eavesdroppers, respectively. For the relay system
\cite{Lee}, according to (\ref{eqn:yit}) and (\ref{eqn:yrt}), with
equal power $P$ allocated to the transmitter and relay, the
achievable rate of the users can be expressed as
\begin{equation}
R_{r}=\log(\det({\boldsymbol I}+\boldsymbol{\Gamma}_{r}^{(t)}  ) ),
\label{eqn:RR1}
\end{equation}
and the $\boldsymbol{\Gamma}_{r}^{(t)}$ according to
(\ref{eqn:SINRrwoc}) is given as
\begin{equation}
\boldsymbol{\Gamma}_{r}^{(t)}
=\sum_{k=1}^{k=K}\frac{P}{N_{k}}{\boldsymbol H}_{kr} {\boldsymbol
H}_{kr}^H({\boldsymbol I}+\frac{P}{N_{t}}{\boldsymbol H}_{i}^{(pt)}
{{\boldsymbol H}_{i}^{(pt)}}^H ), \label{eqn:RR2}
\end{equation}
Similarly, for eavesdropper e the achievable rate assuming global
knowledge for all the links is given by
\begin{equation}
R_{e}=\log(\det({\boldsymbol I}+\boldsymbol{\Gamma}_{e}^{(t)}  ) ),
\label{eqn:RE1}
\end{equation}
and $\boldsymbol{\Gamma}_{e}^{(t)}$ according to
(\ref{eqn:SINRewoc}) is described by
\begin{equation}
\boldsymbol{\Gamma}_{e}^{(t)} = ({\boldsymbol I
+\boldsymbol\varDelta})^{-1}{\frac{P}{N_{t}}{\boldsymbol
H}_{e} {{\boldsymbol H}_{e}}^H}, \label{eqn:RE2}
\end{equation}
where
\begin{equation}
\boldsymbol\varDelta = {
\sum_{e=1}^{N}\sum_{k=1}^{K}\frac{P}{N_{k}}{\boldsymbol H}_{ke} {\boldsymbol
H}_{ke}^H({\boldsymbol I}+\frac{P}{N_{t}}{\boldsymbol H}_{i}^{(pt)}
{{\boldsymbol H}_{i}^{(pt)}}^H)}, \label{eqn:RE21}
\end{equation}

With (\ref{eqn:RR1}) and (\ref{eqn:RE1}) the secrecy rate for the
multiple users is expressed as
\begin{equation}
R=\sum_{r=1}^{T}\sum_{e=1}^N[R_{r}-R_{e}]^{+}, \label{eqn:R}
\end{equation}
where $[x]^{+}=max(0,x)$.

Our objective is to develop an algorithm to select the best set of
relay nodes to perform relay or jammer functions in order to
maximize the secrecy rate. Then, the optimization problem can be
formulated as
\begin{equation}
\begin{aligned}
& \underset{k,i}{\text{max}}
& & R \\
& \text{s.t.} & &  k, i \in \boldsymbol{\Psi}  \\
\end{aligned}
\label{eqn:RF}
\end{equation}
where $\boldsymbol{\Psi}$ represents the relay node poll. The
quantities $k$ and $i$ denote the selected relay and jamming
function nodes, respectively.

\section{Buffer-aided Relay and Jammer Function Selection (BF-RJFS)}

In this section a novel relay and jammer function selection policy
based on the max-ratio relay selection policy in \cite{Gaojie} and
SINR criterion in \cite{Jingchao} is proposed.

\subsection{Motivation}

The aforementioned max-min and max-link relay selection policy are
effective in improving the transmission rate in the relay system,
but the eavesdropper is not taken into consideration. Thus, we apply
the max-ratio relay selection which considers the existence of an
eavesdropper. Unlike the max-ratio relay selection policy, for relay
selection we use the SINR criterion. To further enhance the secrecy
rate performance, we assume the system operates in an opportunistic
scheme so that the relay can also act as a jammer to the
eavesdropper.  
In the following, we will give the details of the proposed
algorithm.

\subsection{Algorithm Description}

We assume that the total number of relay nodes is $Q$. To apply the
opportunistic scheme in the system, an initial state is set in each
relay according to the SINR without jamming:
\begin{equation}
\varGamma_q=\rm{arg} \max_{q \in \boldsymbol{\Psi}}
\det\Big({\boldsymbol{H}_q
{\boldsymbol{H}_q}^{H}}\Big), \label{eqn:varGammam}
\end{equation}
where the jamming relay nodes also transmit the signals to the users, the
selected relay can be either best SINR performance ones which will benefit for
the legitimate users transmission or worst SINR performance ones which provide
jamming to the eavesdropper. Here we choose relays with the best SINR
performance. When the $K$ relays that forward the signals to the users are
determined, the relays used for signal reception are chosen based on the SINR
criterion, as given by
\begin{equation}
\boldsymbol{\phi}_{m}=\rm{arg} \max_{m \in \boldsymbol{\Psi}} \left(
({\boldsymbol{I}+\boldsymbol{\Gamma}_{e}^{(t)}})^{-1}({\boldsymbol{I}+\boldsymbol{\Gamma}_{m}^{(t)}})\right),
\label{eqn:sinrct}
\end{equation}
where $\boldsymbol{\phi}_{m}$ represents the selected relays and $\boldsymbol{\Gamma}_{m}^{(t)}$ is the SINR corresponding to
the $m$th relay which is calculated based on (\ref{eqn:SINRiwoc})
and is given by
\begin{equation}
\boldsymbol{\Gamma}_{m}^{(t)}
=({\boldsymbol{I}+\boldsymbol \varDelta_{m}'})^{-1}({\boldsymbol{H}_{m}\boldsymbol{H}_{m}^{H}}),
\label{eqn:gammam}
\end{equation}
where
\begin{equation}
\boldsymbol \varDelta_{m}'=\sum_{k=1}^{K}\boldsymbol{H}_{km}
\boldsymbol{H}_{m}^{(pt)}{\boldsymbol{H}_{m}^{(pt)}}^{H}\boldsymbol{H}_{km}^{H},
\label{eqn:gammam1}
\end{equation}
with the SINR calculated for the $e$th eavesdropper described by
\begin{equation}
\boldsymbol{\Gamma}_{e}^{(t)}=({\boldsymbol I
+\boldsymbol \varDelta_{e}'})^{-1}({\frac{P}{N_{t}}{\boldsymbol
H}_{e} {{\boldsymbol H}_{e}}^H}),
\label{eqn:gammae}
\end{equation}
where
\begin{equation}
\boldsymbol
\varDelta_{e}'=\sum_{e=1}^{N}\sum_{k=1}^{K}\frac{P}{N_{k}}{\boldsymbol H}_{ke}
{\boldsymbol H}_{ke}^H({\boldsymbol I}+\boldsymbol{\xi}), \label{eqn:gammae1}
\end{equation}
and
\begin{equation}
\boldsymbol{\xi}=\frac{P}{N_{t}}{\boldsymbol H}_{m}^{(pt)}
{{\boldsymbol H}_{m}^{(pt)}}^H, \label{eqn:xi}
\end{equation}
Here the selection of the $T$ relays used for signal reception is
described. The selection of the set of jamming relays is performed
simultaneously. Apart from the $T$ selected relays, the rest of the
relays are selected with the SINR criterion:
\begin{equation}
\boldsymbol{\phi}_{n}= \rm{arg} \max_{n \in \boldsymbol{\Psi}}\left(
({\boldsymbol{I}+\boldsymbol{\Gamma}_{e}^{(t)}})^{-1}({\boldsymbol{I}+\boldsymbol{\Gamma}_{n}^{(t)}})\right),
\label{eqn:sinrct2}
\end{equation}
where $\boldsymbol{\Gamma}_{n}^{(t)}$ is the SINR corresponding to
the $n$th relay which is calculated based on (\ref{eqn:SINRrwoc})
and is given by
\begin{equation}
\boldsymbol{\Gamma}_{n}^{(t)}=
{\boldsymbol{H}_{nr}\boldsymbol{H}_{n}^{(pt)}{\boldsymbol{H}_{n}^{(pt)}}^{H}\boldsymbol{H}_{nr}^{H}},
\label{eqn:gamman}
\end{equation}
With the SINR calculated for the $e$th eavesdropper given by
\begin{equation}
\boldsymbol{\Gamma}_{e}^{(t)}=({\boldsymbol
I +\boldsymbol \varDelta_{e}''})^{-1}({\frac{P}{N_{k}}
\boldsymbol{H}_{ne}\boldsymbol{H}_{n}^{(pt)}{\boldsymbol{H}_{n}^{(pt)}}^{H}\boldsymbol{H}_{ne}^{H}}),
\label{eqn:gammae2}
\end{equation}
where
\begin{equation}
\boldsymbol
\varDelta_{e}''=\sum_{e=1}^{N}\sum_{k=1}^{K}\frac{P}{N_{k}}{\boldsymbol H}_{ke}
{\boldsymbol H}_{ke}^H({\boldsymbol I}+\boldsymbol{\xi}), \label{eqn:gammae22}
\end{equation}
and
\begin{equation}
\boldsymbol{\xi}=\frac{P}{N_{t}}{\boldsymbol H}_{ne} {{\boldsymbol
H}_{ne}}^H, \label{eqn:xi2}
\end{equation}
Then the relays used for jamming in the next time slot are selected.
With a loop of the selection of receiving relays and jamming relays
the system can provide a better secrecy performance as compared to
conventional relay systems. In Algorithm 1 the main steps are
detailed.

{\footnotesize
\begin{algorithm}
\caption{BF-RJFS Algorithm}
\begin{algorithmic}
\FOR {$k=1:K$} \FOR {$q=1:Q$} \STATE $\varGamma_q=\rm{arg} \max_{q \in
\boldsymbol{\Psi}} \det({\boldsymbol{H}_q {\boldsymbol{H}_q}^{H}})$ \ENDFOR
\STATE ${\varGamma_k}^{0}=\varGamma_q$ \STATE $Q^{0}=[1 \quad 2 \quad \cdots
\quad q-1 \quad q+1 \cdots Q]$ \ENDFOR \STATE $Q=Q^{0}$ \LOOP \FOR {$i=1:T$}
\FOR {$m=1:Q^{0}$} \STATE $\boldsymbol{\Gamma}_{m}^{(t)}
=({\boldsymbol{I}+\boldsymbol
\varDelta_{m}'})^{-1}({\boldsymbol{H}_{m}\boldsymbol{H}_{m}^{H}})$ \STATE
$\boldsymbol{\Gamma}_{e}^{(t)}=({\boldsymbol I +\boldsymbol
\varDelta_{e}'})^{-1}({\frac{P}{N_{t}}{\boldsymbol H}_{e} {{\boldsymbol
H}_{e}}^H})$ \STATE $\boldsymbol{\phi}_{m}=\rm{arg} \max_{m \in
\boldsymbol{\Psi}} \left(
({\boldsymbol{I}+\boldsymbol{\Gamma}_{e}^{(t)}})^{-1}({\boldsymbol{I}+\boldsymbol{\Gamma}_{m}^{(t)}})\right)
$ \STATE $Q^{1}=[1 \quad 2 \quad \cdots \quad m-1 \quad m+1 \cdots M]$ \ENDFOR
\ENDFOR \FOR {$k=1:K$} \FOR {$n=1:M^{1}$} \STATE $\boldsymbol{\Gamma}_{n}^{(t)}
={\boldsymbol{H}_{nr}\boldsymbol{H}_{n}^{(pt)}{\boldsymbol{H}_{n}^{(pt)}}^{H}\boldsymbol{H}_{nr}^{H}}$
\STATE $\boldsymbol{\Gamma}_{e}^{(t)}=({\boldsymbol I +\boldsymbol
\varDelta_{e}''})^{-1}({\frac{P}{N_{k}}
\boldsymbol{H}_{ne}\boldsymbol{H}_{n}^{(pt)}{\boldsymbol{H}_{n}^{(pt)}}^{H}\boldsymbol{H}_{ne}^{H}})$
\STATE $\boldsymbol{\phi}_{n}= \rm{arg} \max_{n \in \boldsymbol{\Psi}}\left(
({\boldsymbol{I}+\boldsymbol{\Gamma}_{e}^{(t)}})^{-1}({\boldsymbol{I}+\boldsymbol{\Gamma}_{n}^{(t)}})\right)
$ \STATE $Q^{0}=[1 \quad 2 \quad \cdots \quad n-1 \quad n+1 \cdots M]$ \ENDFOR
\ENDFOR \ENDLOOP
\end{algorithmic}
\end{algorithm}
}\vspace{-0.5cm}

\section{Simulation Results}

In the simulations, a system with $N_{t}=6$ transmit antennas, $M=3$
users and $N=3$ eavesdroppers is considered. The total number of
relays is set to $Q=6$ and among them $T=3$ and $K=3$ relays nodes
are selected. Each user, eavesdropper and relay node is equipped
with $N_{r}=2$, $N_{e}=2$, $N_{i}=2$ and $N_{k}=2$ receive antennas.

\begin{figure}[ht]
\centering
\includegraphics[width=.38\textwidth]{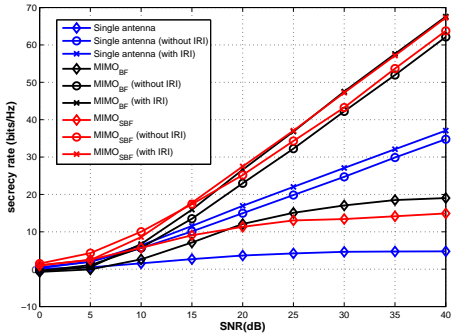}
\caption{Comparison of single-antenna with multi-user MIMO system scenario}
\label{fig:IRI1}
\end{figure}

In Fig. \ref{fig:IRI1}, in a single-antenna scenario, the secrecy
performance with the proposed algorithm is better than that with the
conventional algorithm. With IRI cancellation, the secrecy rate is
better than the one without IRI cancellation. Compared with single
antenna scenario, the multi-user MIMO system contributes to the
improvement in the secrecy rate. In addition, the proposed algorithm
with the selected set of buffer-aided relays (SBF) selection policy
is better than the conventional buffer-aided relay selection (BF).


In Fig. \ref{fig:IRI1}, the power allocation technique is implemented and the
parameter $\eta$ indicates the power allocated to the transmitter. If we assume
in the equal power scenario that the power allocated to the transmitter as well
as relays are both $P$, then the power allocated to the transmitter is $\eta P$
and the power allocated to the relays is $(2-\eta)P$. From Fig. \ref{fig:IRI1}
we can see that with more power allocated to the transmitter the secrecy rate
will decrease. Comparing $(a)$ with $(b)$, when $\eta<1.5$ the secrecy rate
performance in IRI cancellation scenario is better than that without IRI
cancellation. When $\eta>1.5$ and without IRI cancellation, the secrecy rate
can achieve a better performance. \vspace{-0.15cm}

\section{Conclusion}

In this work, we have proposed algorithms to select a set of relay
nodes to enhance the legitimate users' transmission and another set
of relay nodes to perform jamming of the eavesdroppers. The proposed
selection algorithms can exploit the use of the buffers in the relay
nodes that may remain silence during the data transmission.
Simulation results show that the proposed buffer-aided relay and
jammer function selection (BF-RJFS) can provide a better secrecy
rate performance in a multiuser MIMO relay system than existing
buffer-aided relay systems.
\vspace{-0.6cm}


\end{document}